\documentstyle[11pt]{article}
\begin{document}
\begin{titlepage}
\begin{centering}
 
{\ }\vspace{1cm}
 
{\Large\bf Symmetry Tests within the Standard Model}\\
\vspace{0.5cm}
{\Large\bf and Beyond}\\
\vspace{0.5cm}
{\Large\bf from Nuclear Muon Capture}\footnote{To appear in the
Proceedings of the {\sl Monte Verit\`a Workshop on Exotic Atoms,
Molecules and Muon Catalyzed Fusion\/}, July 19-24, 1998,
Ascona, Switzerland}\\
\vspace{1cm}
Jan Govaerts\\
\vspace{1.0cm}
{\em Institut de Physique Nucl\'eaire}\\
{\em Universit\'e catholique de Louvain}\\
{\em B-1348 Louvain-la-Neuve, Belgium}\\
{\tt govaerts@fynu.ucl.ac.be}\\

\vspace{1cm}

\begin{abstract}

\noindent Precision measurements in nuclear muon capture
on the proton and $^3$He allow for tests of the Standard Model
for the strong and electroweak interactions, complementary to those achieved
in high energy experiments. The present situation
and future prospects are reviewed, emphasizing where
renewed efforts could prove to be rewarding in exploring ever further
beyond the confines of the Standard Model.
\end{abstract}

\end{centering} 

\vspace{60pt}

\noindent hep-ph/9809432\\
September 1998

\end{titlepage}

\section{Introduction}

With the availability of intense muon beams of well defined 
characteristics, such as those at the Paul Scherrer Institute,
as well as much improved and new experimental techniques, muonic 
physics has regained much of its past impetus. Precision tests 
of the Standard Model of the strong and electroweak interactions 
have become a reality in recent years, with rare or dominant 
muon decay modes \cite{decay} and nuclear capture processes 
contributing to possibly unvei\-ling the new physics which is
lurking beyond the confines of the Standard Model, in ways 
complementary to those of experiments at much higher energies.

In the cases of nuclear muon capture on $^3$He and the proton, 
achieved \cite{AV,He3} or foreseen~\cite{AV,H}
precisions allow for specific tests of symmetries both of the
hadronic sector of the Standard Model and the electroweak 
interaction \cite{He3,Gov1,Gov2,Gov3}. For instance, muon capture on 
hydrogen within the Standard Model involves directly the nucleon matrix
elements of the vector and axial quark current operators, which,
by virtue of Lorentz covariance, are parametrised according to the
expressions
\begin{equation}
<n|\bar{d}\gamma_\mu u|p>=\bar{n}\,\left[g_V\gamma_\mu+ig_M\sigma_{\mu\nu}
\frac{q^\nu}{2M_N}+g_S\frac{q_\mu}{2M_N}\right]\,p\ \ \ ,
\label{eq:vector}
\end{equation}
\begin{equation}
<n|\bar{d}\gamma_\mu\gamma_5 u|p>=\bar{n}\,\left[g_A\gamma_\mu\gamma_5
+g_P\gamma_5\frac{q_\mu}{m_\mu}+ig_T\sigma_{\mu\nu}\gamma_5
\frac{q^\nu}{2M_N}\right]\,p\ \ \ ,
\label{eq:axial}
\end{equation}
where the quantities $g_V$, $g_M$, $g_S$, $g_A$, $g_P$ and $g_T$ are
form factors which are functions of the momentum transfer invariant $q^2$
with $q^\mu=p^\mu_n-p^\mu_p$ ($u$, $d$, $p$ and $n$ stand for the Dirac
quantum spinor field operators of massive spin 1/2 re\-la\-ti\-vis\-tic 
particles, $M_N$ for the average nucleon mass and $m_\mu$ for the
muon mass). These form factors provide for a phenomenological 
parametrisation of the non perturbative quark bound state structure of 
the nucleons, to be determined from experimental observables and 
symmetry considerations. Requiring invariance of the
above matrix elements under time reversal implies all these
form factors to be real under complex conjugation. Imposing
exact $G$-parity invariance, {\it i.e.\/} exact isospin and charge 
conjugation symmetry, implies that the second-class form factors
$g_S$ and $g_T$ vanish identically for all $q^2$ values (isospin breaking 
effects are such that $|g_S/g_V|$ and $|g_T/g_A|$ are expected \cite{Shiomi}
not to exceed 0.01 to 0.02). Likewise,
in the limit of the exact conservation of the vector current---the CVC
hypothesis---, the remaining vector current form factors $g_V$ and $g_M$
are related to those of the electromagnetic current, which are probed through
electron scattering experiments. From the latter data \cite{Mergell},
one deduces $g_V(q^2_0)=0.9755\pm 0.0005$ and $g_M(q^2_0)=3.582\pm 0.003$,
$q^2_0=-0.88\ m^2_\mu$ being the invariant momentum transfer relevant
to muon capture on the proton. The value for $g_A(q^2_0)$ follows
from $g_A(q^2=0)=1.2601\pm 0.0025$ \cite{PDG} and the nucleon
axial charge radius \cite{Bern1}, so that $g_A(q^2_0)=1.238\pm 0.003$. 
Finally, the value for $g_P(q^2)$ is related to that of $g_A(q^2)$ 
through the partial conservation of the axial current (PCAC) hypothesis, 
which in modern terms is embodied in the approximate chiral symmetries 
of the underlying theory for the strong interactions among quarks, 
namely quantum chromodynamics (QCD). Chiral perturbation theory leads 
to a value for $g_P(q^2_0)$ which depends in particular on the 
pion-nucleon coupling constant. The latest prediction \cite{Bern2} 
is precise to 2.7\%,
\begin{equation}
g_P(q^2_0)=8.44\pm 0.23\ \ \ ,\ \ \ 
\frac{g_P(q^2_0)}{g_A(0)}=6.70\pm 0.18\ \ \ .
\label{eq:theory}
\end{equation}

This result is in fair agreement with the present experimental value
stemming from ordinary muon capture on the proton \cite{Bardin},
$g_P(q^2_0)/g_A(0)=6.9\pm 1.5$, precise to 22\%. However, it is in flagrant
conflict with a recent radiative muon capture measurement \cite{Jonk}
precise to 8\%, namely $g_P(q^2_0)/g_A(0)=9.8\pm 0.8$, which thus disagrees
with the theoretical prediction by a large 4.2 $\sigma$ margin.

Clearly, such a situation in the hadronic sector of the Standard Model
calls for a renewed effort in a precision measurement of an observable
in muon capture on hydrogen which is sensitive to $g_P$, both to reach 
the precision level of the theoretical prediction as an important test of our 
understanding of the chiral symmetry pro\-per\-ties of non perturbative QCD,
as well as to dispell the present conflict within the experimental situation.
As the discussion which is to follow will illustrate, this is but one
instance of a precision measurement in nuclear muon capture which offers
the potential for testing underlying symmetries of the strong and electroweak
interactions.

More specifically, we shall concentrate on three types of observables.
First, a measurement of the statistical capture rate on $^3$He
to the triton channel \cite{AV,He3}, 
$\lambda_{\rm stat}^{\rm exp}=1496\pm 4$ s$^{-1}$, precise to 0.3\%, 
which agrees remarkably well with the theoretical prediction \cite{Jim1} of
$\lambda_{\rm stat}^{\rm theor}=1497\pm 12$ s$^{-1}$. Second, the triton
asymmetry for capture in a polarised $\mu^-\,^3$He system, whose
vector analysing power $A_v$ is predicted \cite{Jim1} to be
$A^{\rm theor}_v=0.524\pm 0.006$, to be compared to the preliminary
experimental value $A^{\rm exp}_v=0.63\pm 0.09^{+0.11}_{-0.14}$ \cite{Souder}.
And third, the foreseen 1\% precision in the measurement of the
singlet capture rate on hydrogen \cite{AV,H}, which should allow for
a determination of $g_P(q^2_0)$ to better than 6\%.
The prospects offered by these different observables will be considered
first within the hadronic sector of the Standard Model, and next, 
within a phenomenological context beyond that Model. The presentation
thus follows that same outline.

\section{Muon Capture within the Standard Model}

Since the $(p,n)$ and ($^3$He,$^3$H) systems are both spin 1/2
isospin doublets, a phenomenological description of muon capture on
either hydrogen or $^3$He proceeds in a similar manner. Thus within
the Standard Model, the effective Hamiltonian for this semi-leptonic 
process reads
\begin{equation}
{\cal H}^L_{\rm eff}=\frac{g^2_L}{8M^2_W}\,V^L_{ud}\,
{J^\mu_{\rm lept}}^\dagger\,{J_{\rm hadr}}_\mu\ \ \ ,\ \ \ 
\frac{g^2_L}{8M^2_W}=\frac{G_F}{\sqrt{2}}\ \ \ .
\label{eq:Heff1}
\end{equation}
Here, $G_F/\sqrt{2}$ represents the Fermi coupling strength,
$V^L_{ud}=0.9751\pm 0.0006$ \cite{PDG}
the Cabibbo-Kobayashi-Maskawa up-down quark flavour mixing 
matrix element, and $J^\mu_{\rm lept}$, $J^\mu_{\rm hadr}$ the 
leptonic and hadronic charged currents, respectively. For the muon 
leptonic flavour, $J^\mu_{\rm lept}=\bar{\mu}\gamma^\mu(1-\gamma_5)\nu_\mu$,
while the hadronic current is of the $(V\!-\!A)$ form,
$J^\mu_{\rm hadr}=V^\mu_{\rm hadr}-A^\mu_{\rm hadr}$, with matrix
elements $V^\mu_{\rm hadr}$ and $A^\mu_{\rm hadr}$ of the quark
vector and axial current operators parametrised as in (\ref{eq:vector})
and (\ref{eq:axial}) (in the case of muon capture on $^3$He,
the relevant form factors are denoted rather as $F_V$, $F_M$, $F_S$,
$F_A$, $F_P$ and $F_T$). This description corresponds to the so called
``elementary particle model" (EPM) approach \cite{Prim}, in which the
underlying bound state structure of nuclei is represented through
phenomenological form factors. For capture on hydrogen, the relevant
momentum transfer is $q^2_0=-0.88\ m^2_\mu$, while for capture on $^3$He,
it is $q^2_1=-0.954\ m^2_\mu$.

Values for these form factors for the $(p,n)$ system have been discussed
above. For the ($^3$He,$^3$H) system, the authors of \cite{Jim1} performed
a very careful assessment of these values, with the following conclusions.
For the vector current, one has the first-class form factors
$F_V(q^2_1)=0.834\pm 0.011$ and $F_M(q^2_1)=-13.969\pm 0.052$.
For the axial current, $F_A(q^2_1)=-1.052\pm[0.005-0.010]$ stems for
the $\beta$-decay rate of $^3$H and an educated guess as to the 
$q^2$-dependence of this form factor, which attempts at including
mesonic exchange current corrections. The uncertainty in this dependency
leads to the range $[0.005-0.010]$ in the error given for $F_A(q^2_1)$,
with the truth lying somewhere in between \cite{Jim1}. Consequently,
uncertainties of results to be quoted hereafter will include this range
of values for $F_A(q^2_1)$. The value for $F_P(q^2_1)$ is again determined
from the PCAC hypothesis, which implies 
$F^{\rm PCAC}_P(q^2)=2m_\mu M\,F_A(q^2)/(m^2_\pi-q^2)$, $M$ being the average
mass value of the initial and final nuclear states (strictly speaking,
this relation assumes that the $q^2$ dependencies of $F_A(q^2)$ and
the $\pi-^3$He-$^3$H coupling constant are identical \cite{Jim1}). 
Finally, in the limit 
of exact $G$-parity invariance, $F_S$ and $F_T$ vanish identically, with 
isospin breaking corrections being at most of a few percent.

Given these values, the statistical capture rate $\lambda_{\rm stat}$ 
on $^3$He to the triton channel, as well as the triton vector analysing 
power $A_v$, are predicted to be \cite{Jim1}
$\lambda^{\rm theor}_{\rm stat}=1497\pm[12-21]$ s$^{-1}$ and
$A^{\rm theor}_v=0.524\pm[0.006-0.006]$, with sensitivities to $F_A$ and
$F_P$ given by $F_A/{\cal O}\,d{\cal O}/dF_A=(1.521,-0.134)$ and
$F_P/{\cal O}\,d{\cal O}/dF_P=(-0.116,-0.377)$ where
${\cal O}=(\lambda_{\rm stat},A_v)$ in the same order.
Note the rather large sensitivity of the capture rate to $F_A$, whose
uncertainty thus dominates that of $\lambda^{\rm theor}_{\rm stat}$, 
a situation which is opposed to that for $A_v$, the latter observable being 
also over three times more sensitive to $F_P$ than is $\lambda_{\rm stat}$. 
Consequently, a combined precision measurement of both
$\lambda_{\rm stat}$ and $A_v$ would enable a model independent determination
of $F_A$ and $F_P$, and thereby a convincing test
of nuclear PCAC. A very precise value for $\lambda_{\rm stat}$ is indeed 
available \cite{He3}, but the preliminary result \cite{Souder} for $A_v$
is not to the required standard.

More specifically, given the result
$\lambda_{\rm stat}^{\rm exp}=1496\pm 4$ s$^{-1}$ \cite{He3}, 
and fixing the values
for all form factors as explained above with the exception of $F_P$,
the nuclear PCAC test is \cite{He3,Gov1,Gov2}
$F_P/F^{\rm PCAC}_P=1.004\pm[0.076-0.132][{\rm exp}:0.023]$, where the first
two numbers in brackets include all theoretical and experimental
uncertainties and correspond to the range of values associated
to the uncertainty in $F_A$, while the number indicated with ``exp" only
includes the uncertainty stemming from the experimental error on
$\lambda_{\rm stat}^{\rm exp}$ alone. Clearly, this test of nuclear PCAC
precise to about 10\% could be improved to some extent were a better value
for $F_A$ to be available independently. Within an impulse 
ap\-pro\-xi\-ma\-tion nuclear model calculation including mesonic exchange 
corrections \cite{Jim2}, the same experimental result leads to a 18\% 
precise PCAC test at the nucleon level, 
$g_P/g^{\rm PCAC}_P=1.05\pm 0.19$ \cite{Jim2}. 
Even though the precision of the theoretical prediction (\ref{eq:theory}) 
is yet to be attained, these conclusions show agreement with QCD chiral 
perturbation theory, as opposed to the result of \cite{Jonk} (incidentally, 
note that the argument may be turned around, and used to determine a rather 
precise value for the $\pi-^3$He$-^3$H coupling constant \cite{Muk}).

Similarly, given the same purpose, let us consider the triton vector 
analysing power $A_v$, assuming a value precise to 1\% centered onto the
theoretical prediction of $A^{\rm theor}_v=0.524$. All other form 
factors being fixed at their specified values, 
the nuclear PCAC test for $F_P$ would then
be precise to 3.9\%, irrespective of the uncertainty on $F_A$ in the
range $[0.005-0.010]$, while including only the experimental error
of 1\% on $A_v$ would provide a PCAC test for $F_P$ to 2.7\%.
It may also be shown that extracting combined values for $F_A$ 
and $F_P$ from both $\lambda_{\rm stat}$ and $A_v$, with a precision 
on $F_A$ at least as good as the present range of $[0.005-0.010]$,
requires a measurement of $A_v$ to at least 1\% relative precision, 
no small feat by any means, but a worthy experimental challenge indeed!

Similar considerations may be developed for the second-class form 
factors $F_S$ and $F_T$, assuming all other form factors set to their 
specified values (in particular, it may be shown that by letting $F_P$ vary
within 10\% of its value, the values for $F_S$ and $F_T$ also vary 
within their respective uncertainties). Here again, it is $A_v$ which
offers the better prospects for improvement, 
with sensitivities such that \cite{Gov1,Gov2,Gov3}
$1/{\cal O}\,d{\cal O}/dF_S=(0.007,0.017)$ and
$1/{\cal O}\,d{\cal O}/dF_T=(-0.006,-0.019)$ for 
${\cal O}=(\lambda_{\rm stat},A_v)$ in the same order.
Specifically, with the result 
$\lambda_{\rm stat}^{\rm exp}=1496\pm 4$~s$^{-1}$, and assuming either 
one of the factors $F_S$ or $F_T$ to vanish in turn, one obtains
$F_S=-0.062\pm[1.18-2.02]\,[{\rm exp}:0.38]$,
$F_T=0.075\pm[1.43-2.45]\,[{\rm exp}:0.46]$. These results improve
on the existing situation for these second-class form factors in the 
impulse approximation \cite{Holstein}, with $g_S=-0.5\pm2.4$ and
$g_T=0.4\pm 2.0$ for capture on $^3$He, and $g_S=-0.4\pm 2.3$ and
$g_T=0.3\pm 1.4$ for capture on hydrogen. On the other hand,
given a measurement of $A_v$ precise to 1\% in the manner assumed above, 
the corresponding uncertainties would be $[0.9-0.9]\,[{\rm exp}:0.58]$
and $[0.8-0.8]\,[{\rm exp}:0.54]$ for $F_S$ and $F_T$, respectively.
Hence here again, one would gain both from a better theoretical knowledge
of $F_A$, as well as from a 1\% precise measurement of $A_v$.

Turning to muon capture on hydrogen, a similar analysis may be applied.
Sensitivities to form factors of the singlet capture rate $\lambda_S$ 
are as follows, $g_X/\lambda_S\,d\lambda_S/dg_X=
(0.47,0.15,1.57,-0.18)$ for $g_X=(g_V,g_M,g_A,g_P)$, and
$1/\lambda_S\,d\lambda_S/dg_X=(0.023,0.024)$ for $g_X=(g_S,g_T)$.
With regards to the first-class form factors, the situation is comparable
to that for $^3$He, with the important difference however, that the value 
for $g_A(q^2_0)$ is known to much better precision. With regards to the
second-class form factors $g_S$ and $g_T$ as well, 
the sensitivity is also much improved.

Specifically, a 1\% precise measurement \cite{AV,H} of $\lambda_S$ centered
at its theoretical value implies 
$g_P=8.44\pm[0.50]\,[{\rm exp}:0.46]$---namely a PCAC test at the
nucleon level precise to 5.9\% (exp: 5.5\%)---, as well as
$g_S=0.0\pm[0.51]\,[{\rm exp}:0.43]$ and
$g_T=0.0\pm[0.50]\,[{\rm exp}:0.42]$, thereby zooming into the theoretically
expected range of values for these form factors, and
much improving the limits on second-class currents in the muon
semi-leptonic sector.

\section{Muon Capture beyond the Standard Model}

Any new physics contribution beyond the Standard Model may 
phenomenologically be parametrised according to the following effective
Hamiltonian \cite{Gov1,Gov2,Gov3}
\begin{equation}
\begin{array}{r l}
{\cal H}_{\rm eff}=\frac{g^2}{8M^2}V_{ud}\,\sum_{\eta_1,\eta_2=+,-}\Bigg[&
(h^V_{\eta_1\eta_2})^*\bar{\nu}_\mu\gamma^\mu(1+\eta_1\gamma_5)\mu\,
\bar{d}\gamma_\mu(1+\eta_2\gamma_5)u\,+\,\\
&+(h^S_{\eta_1\eta_2})^*\bar{\nu}_\mu(1+\eta_1\gamma_5)\mu\,
\bar{d}(1-\eta_2\gamma_5)u\,+\,\\
&+\frac{1}{2}(h^T_{\eta_1\eta_2})^*\bar{\nu}_\mu\sigma^{\mu\nu}
(1+\eta_1\gamma_5)\mu\,
\bar{d}\sigma_{\mu\nu}(1-\eta_2\gamma_5)u\,\Bigg]\ \ \ .
\end{array}
\end{equation}
Here, $g$, $V_{ud}$ and $M^2$ are arbitrary parameters, which in the
Standard Model coincide with the quantities introduced in (\ref{eq:Heff1}),
while the coefficients $h^{S,V,T}_{\eta_1\eta_2}$ are effective couplings 
constants parametrising any possible contribution from physics beyond 
the Standard Model in the charge exchange form, 
the upper index cha\-rac\-te\-ri\-zing the tensor property of
the interaction, and the lower indices $\eta_1$ and $\eta_2$ the $\mu$
and $d$ chiralities, respectively. This effective parametrisation is 
analogous to the by now standard one used for muon decay \cite{PDG} in 
terms of coefficients $g^{S,V,T}_{\eta_1\eta_2}$, while a similar one may 
be considered also for $\beta$-decay processes in terms of coefficients
$f^{S,V,T}_{\eta_1\eta_2}$. Clearly in the Standard Model, all these 
coefficients vanish identically, except for $h^V_{LL}\equiv h^V_{--}=1$.
Given such general parametrisations, the effective values for $g^2/8M^2$ and
$V_{ud}$ have to be determined accordingly from the muon decay rate
and the $0^+-0^+$ superallowed $\beta$-decay rates, respectively.

The above interactions contribute to muon capture through the hadronic
matrix elements of the corresponding quark operators. For the vector
and axial currents, the parametrisation in terms of form factors has
been introduced in (\ref{eq:vector}) and (\ref{eq:axial}). Likewise
for the scalar, pseudoscalar and tensor operators $\bar{d}u$,
$\bar{d}\gamma_5 u$ and $\bar{d}\sigma_{\mu\nu} u$, respectively,
the associated nuclear matrix elements may be parametrised in terms of
form factors $g^0_S$, $g^0_P$ and $g^0_T$, or $G_S$, $G_P$ and $G_T$,
for the $(p,n)$ and the ($^3$He,$^3$H) systems respectively,
when ignoring possible recoil order corrections which 
are subdominant in any case.

Sensitivities of the statistical capture rate 
$\lambda_{\rm stat}$ in the case of $^3$He, and of the singlet one 
$\lambda_S$ in the case of hydrogen, to the coefficients 
$h^{S,V,T}_{\eta_1\eta_2}$ are as follows (assuming
that all form factors just introduced, be\-yond the vector and axial current
ones, are set to the value unity, and also that all the 
$h^{S,V,T}_{\eta_1\eta_2}$ coefficients are
real under complex conjugation, and thus do not
lead to potential new sources of CP violation). One has
$1/\lambda_{\rm stat}\,d\lambda_{\rm stat}/dh^X=(2.0,-0.81,0.38,-0.0056,5.82)$
as well as
$1/\lambda_S\,d\lambda_S/dh^X=(2.0,-0.76,0.41, 0.022,-5.23)$, with
$h^X=(h^V_{--},h^V_{-+},h^S_+,h^P_+,h^T_{+-}/2)$ in the same order,
and the definitions $h^S_+=h^S_{++}+h^S_{+-}$ and
$h^P_+=h^S_{++}-h^S_{+-}$. Hence, these sensitivities are comparable
in both cases, except for a possible pseudoscalar interaction.

More explicitly, consider the case of $^3$He, with
$\lambda_{\rm stat}^{\rm exp}=1496\pm 4$ s$^{-1}$ \cite{He3}.
Assuming that only $h^V_{LL}$ is induced with a value different from
unity, as well as $f^V_{LL}$ in the electronic sector, one establishes the
$e\!-\!\mu$ universality test 
\begin{equation}
\begin{array}{r c l}
|h^V_{LL}/f^V_{LL}|^2&=&0.9996\pm[0.0083-0.0140]\,[{\rm exp}:0.0023]\ \ \ ,
\\ \\
|h^V_{LL}/f^V_{LL}|&=&0.9998\pm[0.0042-0.0071]\,[{\rm exp}:0.0013]\ \ \ ,
\end{array}
\label{eq:He3univ}
\end{equation}
to be compared to the usual $e\!-\!\mu$ universality test from $\pi$ decay,
$|h^V_{LL}/f^V_{LL}|^2=1.0040\pm 0.0033$ \cite{Her}. Here again, were the
value of $F_A$ to be improved, a genuinely significant independent test
of $e\!-\!\mu$ universality would become feasible. Otherwise, assuming now
that $h^V_{LL}=1$ and that only one new effective coupling 
$h^{S,V,T}_{\eta_1\eta_2}$ takes a non vanishing value, one infers the
following constraints, 
\begin{equation}
\begin{array}{r c l}
h^V_{-+}&=&0.0005\pm[0.0102-0.0176]\,[{\rm exp}:0.0033]\ \ \ ,\\ \\
h^S_+G_S&=&-0.0012\pm[0.022-0.038]\,[{\rm exp}:0.0071]\ \ \ ,\\ \\
h^P_+G_P&=&-0.078\pm[1.49-2.56]\,[{\rm exp}:0.48]\ \ \ ,\\ \\
\frac{1}{2}h^T_{+-}G_T&=&-0.00008\pm[0.00143-0.00245]\,[{\rm exp}:0.00046]
\ \ \ .
\end{array}
\label{eq:He3constr}
\end{equation}
In particular, the constraints on the scalar $h^S_+$ and tensor
$h^T_{+-}$ interactions are very stringent, and provide a genuine
improvement on the existing situation by a large margin, also when
compared with the electronic sector stemming from $\beta$-decay.
In addition, here again, it may be shown that when $F_P(q^2_1)$ is left 
to vary within 10\% of its value, the above results are in fact quite
robust, since they remain within their
uncertainties. Also note that these limits once again would gain
from a better knowledge of $F_A(q^2_1)$.

The above constraints are valid quite independently of any model
for physics beyond the Standard Model. Nevertheless, it proves useful
to also consider specific model extensions to assess in clearer physical
terms the reach of these limits. Thus for example, within the context
of so called left-right sym\-me\-tric 
SU(2)$_L\times$SU(2)$_R\times$U(1)$_{B-L}$ gauge models \cite{Moh},
the constraint on $h^V_{-+}$ implies $g_R/g_L\,{\rm Re}
\left(e^{i\omega}V^R_{ud}/V^L_{ud}\right)\,\tan\zeta=
-0.0005\pm[0.0102-0.0176]\,[{\rm exp}:0.0033]$, where $g_{R,L}$ and
$V^{R,L}_{ud}$ are gauge coupling constants and Cabibbo-Kobayashi-Maskawa
matrix elements associated to the sectors of right- and left-handed
chiralities, while $\zeta$ is the mixing angle for massive charged
gauge bosons and $\omega$ is a CP vio\-la\-ting phase also following from 
the diagonalisation of the charged gauge boson mass matrix. In particular
for so called manifestly left-right symmetric models with 
$g_R=g_L$, $V^R_{ud}=V^L_{ud}$
and $\omega=0$, the ensuing constraint on the mixing angle $\zeta$ is
competitive with limits stemming from $\beta$-decay \cite{PDG}, and
would also gain from an improvement on $F_A(q^2_1)$.

Another quite popular model extension of the Standard Model are
so called contact interactions, in which a specific energy scale 
$\Lambda$ is associated to a possible substructure of quarks and 
leptons \cite{PDG}. By way of example, and u\-sing the customary 
parametrisation of contact interactions specified in \cite{PDG}, 
the value for $\lambda_{\rm stat}^{\rm exp}$ tran\-slates 
for instance into the following limits for such a compositeness scale,
$\Lambda^V_{-+}>[4.9-3.8]\,[{\rm exp}:8.4]$ TeV (90\% C.L.) as well as
$\Lambda^T_{+-}>[9.3-7.2]\,[{\rm exp}:15.9]$~TeV (90\% C.L.)
(the latter values assume $G_T=1$).
These constraints, of application to interactions coupling the second
leptonic generation to the first quark generation, are complementary
to existing ones \cite{PDG}, which often involve rather the first generation
leptons. In fact, these limits are a genuine competition for high energy 
experiments in the case of charged current electroweak contact interactions, 
for which the energy scale $\Lambda$ is typically in the 2-4~TeV range.

Yet another model extension of the Standard Model are so called
lepto-quark interactions \cite{lepto1,lepto2} (for which the notations
of \cite{lepto2} will be used to refer to leptoquarks and their
Yukawa couplings). Note however that the constraints to be given presently
apply to Yukawa couplings and leptoquark masses coupling the second lepton
generation to the first quark generation. Focusing again onto the more
competitive limits, the scalar and tensor effective 
interactions provide for the followings constraints. In the scalar case,
the combination
\begin{equation}
\left|\frac{\lambda^L_{S_0}\lambda^R_{S_0}}{M^2(S_0)}\,+\,
\frac{\lambda^L_{S_{1/2}}\lambda^R_{S_{1/2}}}{M^2(S_{1/2})}\,+\,
4\frac{\lambda^L_{V_0}\lambda^R_{V_0}}{M^2(V_0)}\,+\,
4\frac{\lambda^L_{V_{1/2}}\lambda^R_{V_{1/2}}}{M^2(V_{1/2})}\right|\ \ \ ,
\end{equation}
is bounded above by $[2.2-3.8]\,[{\rm exp}:0.77]$ TeV$^{-2}$ (90\% C.L.), 
or equivalently by
\begin{equation}
\left[\frac{0.023}{(100\ {\rm GeV})^2}-\frac{0.038}{(100\ {\rm GeV})^2}\right]
\,\left[{\rm exp}:\frac{0.008}{(100\ {\rm GeV})^2}\right]\ \ \ 
(90\%\ {\rm C.L.})\ \ \ ,
\end{equation}
when normalising the leptoquark mass scale to 100 GeV/c$^2$.
Similarly in the tensor case, the combination
\begin{equation}
\left|\frac{\lambda^L_{S_0}\lambda^R_{S_0}}{M^2(S_0)}\,-\,
\frac{\lambda^L_{S_{1/2}}\lambda^R_{S_{1/2}}}{M^2(S_{1/2}(-2/3))}\right|\,
|G_T|\ \ \ ,
\end{equation}
is bounded above by $[0.29-0.50]\,[{\rm exp}:0.10]$ TeV$^{-2}$ (90\% C.L.),
or equivalently by
\begin{equation}
\left[\frac{0.003}{(100\ {\rm GeV})^2}-\frac{0.005}{(100\ {\rm GeV})^2}\right]
\,\left[{\rm exp}:\frac{0.001}{(100\ {\rm GeV})^2}\right]\ \ \  
(90\%\ {\rm C.L.})\ \ \ .
\end{equation}
These limits are quite competitive with, and complementary to existing ones
from high energy experiments \cite{lepto2,PDG}, especially for the last set 
of constraints stemming from tensor type effective interactions.
Once more, note how these results would also gain from an improved knowledge 
of $F_A(q^2_1)$.

The potential for similar tests of physics beyond the Standard Model
from a precision measurement of the singlet capture rate $\lambda_S$
on hydrogen is as follows. Assuming again a 1\% precise result centered
onto the theoretically expected value, the corresponding
constraints are (taking $g^0_S$, $g^0_P$ and $g^0_T$ all equal to unity),
\begin{equation}
\begin{array}{r c l}
h^V_{--}&=&1.0\pm0.0060\,[{\rm exp}:0.0050]\ \ \ ,\\ \\
h^V_{-+}&=&0.0\pm0.0156\,[{\rm exp}:0.0131]\ \ \ ,\\ \\
h^S_+&=&0.0\pm0.0287\,[{\rm exp}:0.0242]\ \ \ ,\\ \\
h^P_+&=&0.0\pm0.55\,[{\rm exp}:0.46]\ \ \ ,\\ \\
\frac{1}{2}h^T_{+-}&=&0.0\pm0.0023\,[{\rm exp}:0.0019]\ \ \ ,
\end{array}
\end{equation}
to be compared to those given in (\ref{eq:He3univ}) and (\ref{eq:He3constr}).
Thus also for muon capture on hydrogen, it is the tensor effective
coupling coefficient $h^T_{+-}$ which would be subjected to the most
stringent constraint, while those for the vector $h^V_{-\pm}$ and
scalar $h^S_+$ coefficients remain also of much interest.

\section{Conclusions}

As this contribution has demonstrated, precision measurements in
muon capture on hydrogen and $^3$He provide for important symmetry
tests of the Standard Model---both in its hadronic as well as in its
electroweak sector---which are competitive with, and complementary to 
experiments at high energies.

The potential for such tests has already been established for the
statistical capture rate on $^3$He to the triton channel, given the
recent 0.3\% precise measurement of \cite{He3}. The physics reach 
of the ensuing contraints could be improved still further through 
a better knowledge of the nuclear axial form factor $F_A(q^2_1)$, 
a problem for which a chiral perturbation approach 
{\it at the nuclear level\/}, including
isospin breaking effects, could be envisaged.

Further progress is to be made through a forthcoming measurement of the 
singlet cap\-ture rate $\lambda_S$ on hydrogen, to a precision better than
1\% \cite{H}. The physics reach of such result is complementary to that 
for $^3$He, since it is much less affected by theoretical uncertainties,
while on the other hand a 0.5\% precision measurement now seems 
feasible \cite{AV}, thereby improving by almost a factor two the 
uncertainties on the symmetry tests discussed in this contribution. 
In particular, this would bring the uncertainty of an experimental test 
of the chiral symmetry prediction for $g_P$ down to the same level of 
precision as the theoretical value \cite{Bern2}.

Finally, renewed efforts in a precision measurement of the vector analysing 
power for the triton asymmetry in capture on $^3$He \cite{Souder}---or 
any other polarisation observable for that matter---, would provide 
for additional stringent symmetry tests as well \cite{Gov3}, 
which would be complementary 
to those stemming from the capture rate result \cite{He3} and be more 
independent of model assumptions. The required precisions are quite 
demanding however, but the experimental challenge is certainly to 
the standard of its possible physics rewards.

\section*{Acknowledgements}

The author wishes to thank Prof. Claude Petitjean for his very kind
invitation to this Workshop, him and all the organisers for a most 
instructive and enjoyable Workshop, and all participants for their interest 
and for their contributions in many ways. Prof. Jules Deutsch is 
also gratefully acknowledged for numerous discussions over the years.


%
\end{document}